# $MoS_2$ Impurities: Chemical Identification and Spatial Resolution of Bismuth Impurities in Geological Material

Maria Gabriela Sales[a], Lucas Herweyer[a], Elizabeth Opila[a], Stephen McDonnell[a,*]

[a]Department of Materials Science and Engineering, University of Virginia, Charlottesville, VA 22904

*Corresponding author: mcdonnell@virginia.edu

**Abstract**
Molybdenum disulfide ($MoS_2$) is the most widely studied transition metal dichalcogenide (TMDC) material, in part because it is a natural crystal present in the earth, thus making it abundant and easily accessible. Geological $MoS_2$ has been used in various studies that look at incorporating $MoS_2$ into devices for nanoelectronics and optoelectronics. However, variations in the electronic properties of a single $MoS_2$ surface are known to exist due to defects that are intrinsic to natural $MoS_2$. This work reports the presence of bismuth impurities in $MoS_2$ with concentrations high enough to be detected by X-ray photoelectron spectroscopy (XPS). These concentrations are further corroborated with inductively coupled plasma optical emission spectroscopy (ICP-OES). Localization of these bismuth clusters is shown using XPS-mapping, and the cluster size is determined to be on the order of tens of microns. This work provides important insights into the nature of impurities that are known to exist in $MoS_2$.

1. **Introduction**

Since the demonstration of the properties of isolated graphene[1], similarly structured 2D materials have been under investigation for their interesting and unique properties. Of particular interest are the transition metal dichalcogenides (TMDCs) family of materials, which, for specific metal and chalcogen combinations, can be semiconducting materials whose band gaps are tunable with thickness, making them a versatile material with a wide range of possible applications. By far the most widely studied TMDC is molybdenum disulfide or $MoS_2$. Like graphene/graphite, $MoS_2$ is easily accessible because it is a naturally occurring compound and commercially available. Several studies have reported the use of geological $MoS_2$ flakes in devices for applications in nanoelectronics[2-4], optoelectronics[5-7], batteries[8], biosensors[9-10], and hydrogen production[11-13], among others.

Geological $MoS_2$ is known to have various intrinsic defects which have been explored for several years. A known issue with geological $MoS_2$ material is its variability in doping across at even tens of nanometers of spatial resolution, which is attributed to defects in the $MoS_2$ crystal[14]. Variability in the doping and electronic properties of the starting $MoS_2$ is attributed to structural and chemical non-uniformities that are a characteristic of geological material. Such variations may greatly affect the properties of any device into which the material is integrated. A prior study observed a correlation between n-type regions and S vacancies or metallic Mo-like clusters, while p-type regions were correlated with regions of high structural defect density in the crystal[14]. Related to this doping variability, ring-like structures a few nanometers in diameter have been found in scanning tunneling microscopy (STM) studies of geological $MoS_2$, and these have been attributed to the electronic effect of the presence of impurity point defects acting as dopants in $MoS_2$[15-16]. A separate study[17] reported the various defects experimentally observed on the surface of geological $MoS_2$ with STM, and these include pits, step edges, voids underneath the top surface, local surface reconstructions, and point defects like vacancies and impurities.

Through various mass spectroscopy techniques, previous studies have shown the presence of impurity elements in natural $MoS_2$ material[15-18]. Addou et al. report bismuth to be the most abundant trace element in natural $MoS_2$, at ~20 parts per million by weight (ppmw)[18]. In the present study, it is reported that Bi is sometimes observed in geological $MoS_2$ with X-ray photoelectron spectroscopy (XPS), indicating that there are localized regions in which the Bi concentrations are higher than ~0.1%, which is the typical detection limit of XPS[19]. The chemical state of these Bi impurities is determined. Estimates of the impurity size and distribution across the MoS$_2$ surface are also reported.

## 2. Material and Methods

Geological MoS$_2$ crystal (Graphene Supermarket[20]) was cleaned by mechanical exfoliation with scotch tape which removed the top most layers of MoS$_2$, leaving a freshly exposed crystal surface. These samples were then loaded into ultra-high vacuum (UHV) for X-ray photoelectron spectroscopy (XPS) in less than 30 minutes. XPS spectra were acquired using Al Kα X-rays (1468.7 eV) at a pass energy of 26 eV in a PHI VersaProbe III system. This XPS tool is also equipped with a scanning X-ray induced (SXI) secondary electron imaging feature for accurate point selection in the micrometer range, and a variety of X-ray spot sizes ranging from 9 to 200 μm. Spectra of geological MoS$_2$ from another vendor (SPI Supplies[21]) are shown in Supplementary Material to confirm that Bi is detected from more than one material source, but all XPS analyses shown in the main text are of MoS$_2$ from Graphene Supermarket.

XPS mapping was performed by selecting a random area of the crystal and through the SXI imaging mode, setting up an array of points that are 100 μm apart horizontally and vertically. The Mo 3$d$ and S 2$p$ spectra were taken at each of these points using an X-ray beam spot size of 100 μm. The points analyzed were within an 800 μm x 800 μm area, for a total of 81 spots checked. These spectra were batch processed using the PHI MultiPak software in order to identify which of the 81 points had Bi peaks present from XPS.

A particle/cluster size check of the bismuth impurities was achieved by centering the X-ray spot on a Bi impurity, and without moving the sample, analyzing the same spot using different X-ray beam spot sizes: 9 μm, 15 μm, 20 μm, 50 μm, 100 μm, and 200 μm. All XPS peak fitting in this work to obtain areas of the Mo 3$d$, S 2$p$, and Bi 4$f$ features was carried out using KolXPD software[22]. All peaks were fit with Voigt lineshapes.

Inductively coupled plasma optical emission spectroscopy (ICP-OES) was performed to support the XPS measurements. A 7 mg geological MoS$_2$ flake (SPI Supplies) was placed inside a high-purity polypropylene centrifuge (VWR International) tube filled with a 13 mL H$_2$O and 2 mL HNO$_3$ acid containing solution. Digestion of MoS$_2$ was performed for several days in a 40 °C water bath with intermittent sonication. Upon completion of digestion, 3-4 emission lines each from Bi, W, Ag, Cd, Fe Re and Ca were detected using ICP-OES (Duo Spectrometer iCAP 6200, ThermoFisher Scientific). The reported concentrations (in ppmw) are averages from all emission lines for a given element. The complete results of all elements detected from ICP-OES are detailed in Supplementary Material.

## 3. Results and Discussion

### 3.1. XPS of MoS$_2$ with Bismuth

Shown in Fig. 1 is a typical XPS spectra of MoS$_2$ with the additional peaks corresponding to Bi impurities in the geological crystal. Using a micrometer-sized X-ray beam spot, peaks that are not attributed to MoS$_2$ were sporadically found at approximately 159.5 eV and 164.8 eV, which correspond to Bi 4$f_{7/2}$ and Bi 4$f_{5/2}$, respectively. Elemental Bi 4$f$ peaks are expected at around 157 eV and 162 eV, which are more than 2 eV lower. This suggests that the Bi impurities that have been known to exist in geological MoS$_2$, as demonstrated in previous studies[18], are not metallic Bi impurities. Based on the peak positions in our XPS analyses, it can be confirmed that these Bi impurities exist in the MoS$_2$ as bismuth oxide or bismuth sulfate.

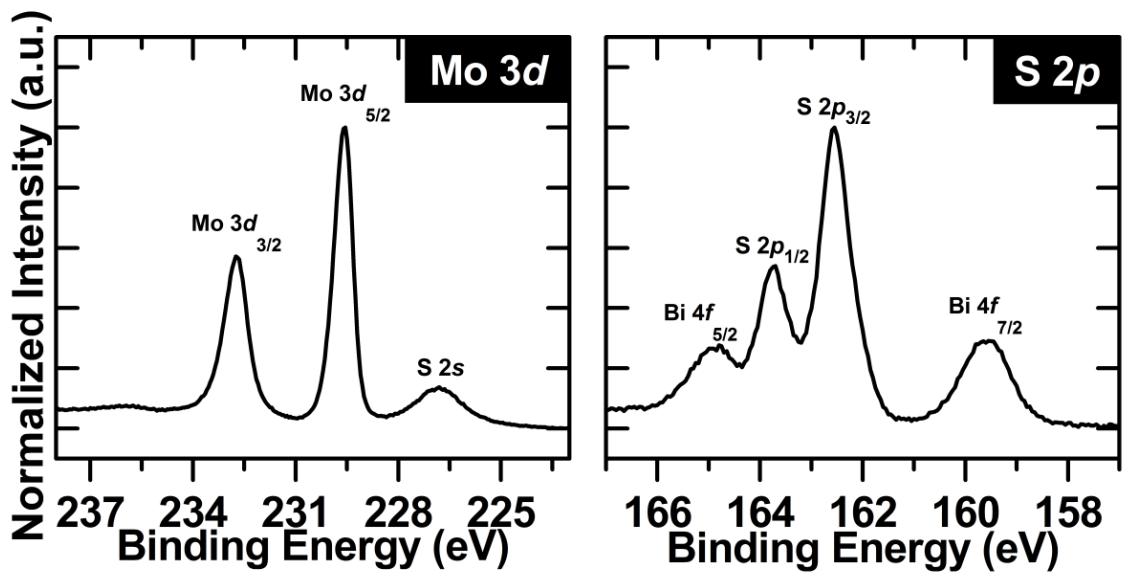

**Figure 1.** Mo 3$d$ and S 2$p$ regions of a geological MoS$_2$ crystal showing additional peaks corresponding to Bi. Spectra taken with a 100 μm X-ray beam spot size.

### 3.2. XPS Mapping

The Mo 3$d$ and S 2$p$ regions were taken at 81 spots in an 800 μm x 800 μm XPS map set up, as presented in Fig. 2a. The results from this 800 μm x 800 μm area is assumed to be representative of the entire MoS$_2$ crystal because this area was chosen at random in the middle of a region where the MoS$_2$ appeared to be smooth. The distribution of bismuth found through XPS mapping reveals that these Bi impurity compounds are present at random localized spots of the bulk geological MoS$_2$ crystal. A map of Bi 4$f_{7/2}$ intensities normalized to the S 2$p_{3/2}$ peak found through XPS is shown in Fig. 2b, which clearly exhibits how localized these Bi impurities are. Out of the 81 analyzed spots, 12 were seen to have a Bi 4$f$ feature, the majority of which are localized in regions which are a few hundred microns across. The normalized spectra of the 12 points with detectable Bi presence are shown in Fig. 2c. It can be seen that their Mo 3$d$ spectra are almost the same except for some peak shifting which is likely due to variable doping[14]. However, major differences are evident in the S 2$p$ spectra because of varying intensities of the Bi 4$f$ features. The spectrum with the highest Bi feature relative to the S 2$p$ is shown as the blue curve, while the one with the lowest non-zero Bi 4$f$ peak area is the red curve. The normalized Bi intensity of the blue curve is 67 times greater than that of the red curve.

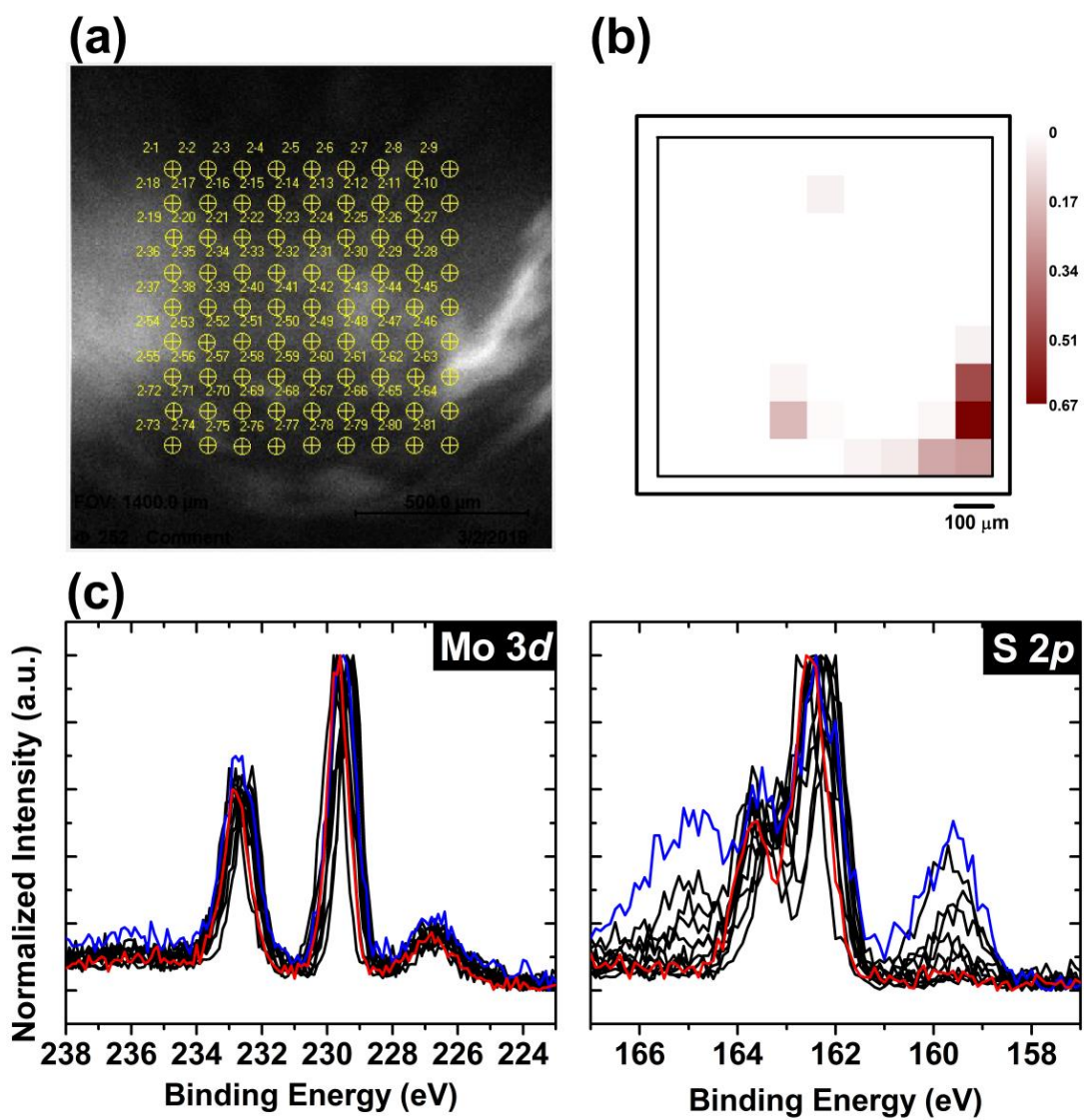

**Figure 2.** (a) Set-up of the array of 81 points spaced 100 µm apart for XPS mapping on an 800 µm x 800 µm surface of geological $MoS_2$. (b) Map of normalized Bi intensities obtained through XPS mapping. (c) Normalized spectra of the 12 analysis spots that were found to have some amount of Bi.

It is worth noting that Bi concentration calculated using relative sensitivity factors would not be appropriate here. This is because concentrations calculated using relative sensitivity factors assume a homogenous distribution. As will be discussed later in this paper, it is likely that all of the Bi clusters observed in this work are actually covered with layers of $MoS_2$, therefore their intensities are attenuated. Simply averaging all intensities and applying sensitivity factors would result in a gross underestimation of the Bi concentration. However, concentrations calculated from each individual spot in the XPS map can still give a sense of the amount of Bi detected in the area analyzed. Details of this calculation are discussed in Supplementary

Material. The average concentration of the Bi signal detected across all 81 spots in the map would give an estimated Bi concentration of 0.17% in the surface area analyzed. This average concentration would explain why Bi is not typically detected in studies which show XPS spectra of geological $MoS_2$[5, 9, 12, 14, 17-18, 23-24]. The concentration is close to the typical limit of detection of XPS, ~0.1%[19].

### 3.3. Cluster Size Determination

It was found that regions with Bi in geological $MoS_2$ are on the order of hundreds of microns, however we note that this does not necessarily mean that this is the size of a single impurity. This region could be composed of clusters of Bi impurity particles which would be indistinguishable from a single large inclusion from our measurements. We are able to estimate the size of the clusters by scanning the same spot with varying X-ray beam spot sizes from 200 to 9 µm. As we vary the spot size from large to small we expect to see changes in the Bi:S ratio as the percentage of the area being dominated by Bi changes. We expect one of three possible changes in this ratio. 1) The Bi:S ratio increases to a point where the S signal becomes zero. This would be interpreted as a Bi cluster on top of $MoS_2$. The cluster size in this case would be bigger than the X-ray spot size at the point when the $MoS_2$ signal dropped to zero. Therefore, the cluster size would be estimated to be between that of the X-ray beam spot sizes that could detect and could not detect S. 2) The Bi:S ratio increases to a point and plateaus. This would be interpreted as a Bi cluster that is bigger than the spot size at the point when the ratio plateaus, but with the cluster being below some layers of $MoS_2$ (hence the Mo and S signals never decay to zero). 3) No change in the Bi:S ratio. This would be determined as a cluster that is bigger than 200 µm, but still below $MoS_2$ layers.

The normalized Mo 3*d* and S 2*p* spectra of two different spots in the $MoS_2$ that have Bi impurities are shown in Fig. 3 below. It is evident in Fig. 3a that the Bi 4*f* features at small spot sizes have comparable intensities, and above a critical spot size, the peak intensity of the Bi 4*f* feature drops relative to the S 2*p* peak.

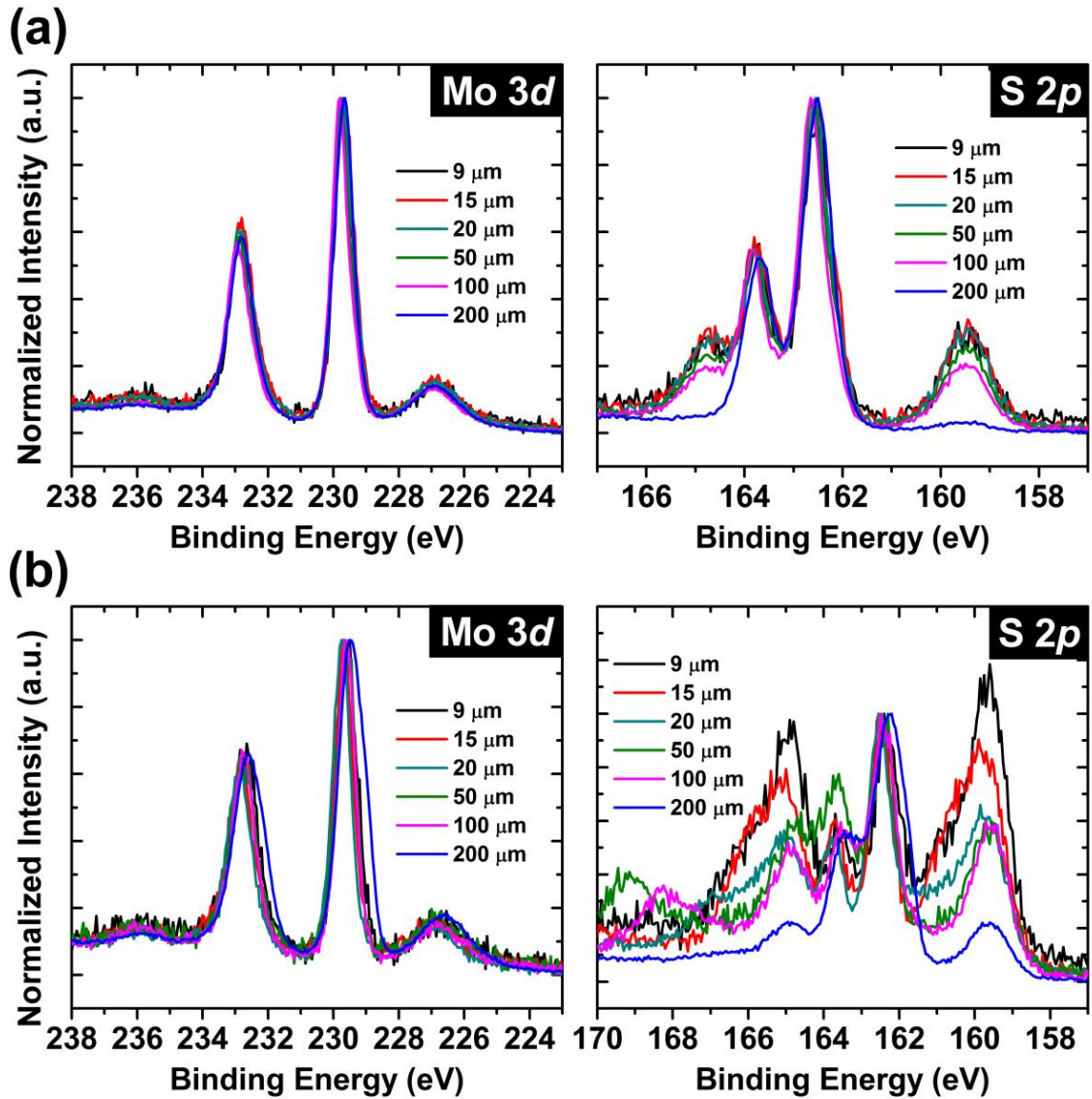

**Figure 3.** Normalized spectra of analysis spot 1 (a) and analysis spot 2 (b), taken with varying X-ray spot sizes.

For Fig. 3b, which is a second analysis spot, the Bi feature is composed of two chemical states, one of the typical Bi 4$f$ peak positions present in MoS$_2$ as shown in Fig. 1, and a second bismuth peak at higher binding energy. A deconvolution of the spectrum taken with the 9 μm X-ray beam spot is shown in Fig. 4, wherein the presence of two Bi oxidation states, Bi$^{3+}$ and Bi$^{5+}$, is exhibited. The peaks for Bi$^{3+}$ are at 159.6 eV and 164.9 eV, while the Bi$^{5+}$ peaks are at 160.8 eV and 166.1 eV, which agree with peak assignments of bismuth oxidation states in other reports[25-26]. Bi$_2$O$_3$ (Bi$^{3+}$) is the most thermodynamically stable form of bismuth oxide, but Bi$_2$O$_5$ (Bi$^{5+}$) and other oxidation states, Bi$^{2+}$ and Bi$^{4+}$, can form from the stable Bi$_2$O$_3$ phase when it is heated to 1100-1300 K[27]. Some of the spectra in the S 2$p$ region of Fig. 3b also show extra features at around 168 to 169 eV, which indicates the presence of some sulfate. Because $Bi_2O_3$ and $Bi_2(SO_4)_3$ both have an oxidation state of $Bi^{3+}$, their Bi peaks would appear at similar binding energies[28]. The presence of sulfate peaks in some of our spectra indicate the possibility that some of the bismuth impurities may exist as $Bi_2(SO_4)_3$ compounds.

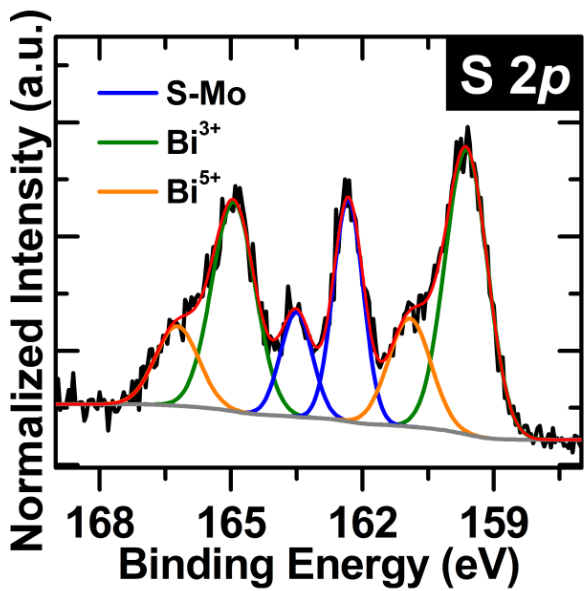

**Figure 4.** Deconvolution of the S 2*p* region at analysis spot 2 taken with a 9 μm X-ray beam spot.

Table 1 shows the normalized Bi intensities at these different spot sizes, which are also plotted in Fig. 5. The Bi intensities for analysis spot 1 are the normalized intensity of the single Bi peak corresponding to $Bi^{3+}$, while for analysis spot 2, the reported Bi intensities are the sum of the normalized intensities of the $Bi^{3+}$ and $Bi^{5+}$ states. We assume that the $Bi^{5+}$ was chemically transformed from $Bi^{3+}$, and thus assign both oxidation states to be part of the same impurity cluster. Table 1 and Fig. 5 clearly show that the Bi signal stays about the same at the smaller spot sizes. At the spot sizes where the normalized Bi intensities plateau, the X-ray beam spot size must be smaller than the impurity cluster size. As the spot size gets bigger, the area that is probed is bigger, and at a critical spot size this area probed becomes bigger than the impurity cluster size. This is seen as a drop in the normalized Bi intensity because there is more $MoS_2$ that contributes to the signal. For analysis spot 1, the first significant drop in the Bi signal is between 20 and 50 μm, while for analysis spot 2 the first significant drop is between 15 and 20 μm. From this, it can be implied that the size of the impurity analyzed in spot 1 is somewhere between 20 and 50 μm. For analysis spot 2, the impurity size is somewhere between 15 and 20 μm.

**Table 1.** Normalized Bi intensities found using different X-ray beam spot sizes at analysis spots 1 and 2.

| X-ray Beam Spot Size | Normalized Bi Intensity | |
|---|---|---|
| | Analysis Spot 1 | Analysis Spot 2 |

| 9 µm    | 0.27 | 1.67 |
| 15 µm   | 0.31 | 1.67 |
| 20 µm   | 0.30 | 1.25 |
| 50 µm   | 0.24 | 0.82 |
| 100 µm  | 0.20 | 0.86 |
| 200 µm  | 0.02 | 0.32 |

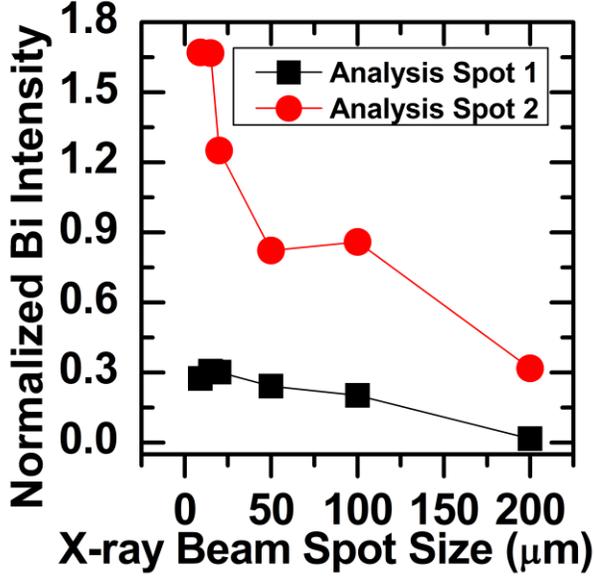

**Figure 5.** Normalized Bi intensities of analysis spots 1 and 2 plotted against X-ray beam spot size.

A key point is that at low spot sizes, while the Bi intensity has increased and plateaued, the detection of MoS$_2$ has never dropped to zero. Assuming approximately spherical clusters, the > 10 µm diameter can be estimated to imply a > 10 µm depth. This is >> the ~10 nm sampling depth of XPS, and so the MoS$_2$ would easily be entirely attenuated by a cluster that was on top of the MoS$_2$. Therefore, we speculate that the observed clusters are actually under at least a few layers of MoS$_2$. We note that the extreme surface sensitivity of XPS means that clusters under even 10 layers of MoS$_2$ are likely not detected at all.

It is possible to estimate the thickness of the MoS$_2$ covering the Bi cluster. This was achieved by comparing the Mo $3d_{5/2}$ intensities of analysis spots 1 and 2 with the same core level intensities for pure geological MoS$_2$ (i.e. no Bi) scanned using the same XPS settings. The integrated XPS intensity, I, of a layer of certain thickness d is given by Eq. 1[29].

$$I = I_\infty [1 - \exp\left(-\frac{d}{\ell \cos\theta}\right)] \quad (1)$$

In this equation, I is the measured intensity of the overlayer, I$_\infty$ is the intensity of a pure, clean, bulk sample of the same material as the overlayer, d is the overlayer thickness, $\ell$ is the effective attenuation length of a photoelectron traveling through the overlayer, and θ is the photoelectron take-off angle measured with respect to the surface normal. Manipulating Eq. 1, the thickness of the overlayer, d, is given by Eq. 2.[29]

$$d = -\ell \cos\theta \ln(\frac{I_\infty - I}{I_\infty}) \quad (2)$$

For our purposes, the core level that we use for these equations is Mo $3d_{5/2}$. The effective attenuation length, $\ell$, of Mo $3d_{5/2}$ in $MoS_2$, which is our overlayer, is taken to be 23.815 Å[30]. The photoelectron take-off angle, θ, is 45°, as given by the geometry of the XPS system used. The $I_\infty$ used was the Mo $3d_{5/2}$ peak intensity from pure $MoS_2$, while I was the Mo $3d_{5/2}$ peak intensity at analysis spot 1 or 2. The peak intensity at analysis spot 1 or 2 is assumed to be the intensity of the overlayer whose thickness we are calculating. The intensities that we used for our calculations were the intensities at 9 µm, which is the smallest X-ray spot size, in order to ensure that the X-ray spot is not bigger than the Bi impurity. This is important so that it is safe to assume that all of the Mo and S signals are from some thickness, d, of $MoS_2$ on top of the Bi impurity. The calculated $MoS_2$ overlayer thickness was 3.17 nm for analysis spot 1, and 0.85 nm for analysis spot 2. The thicker overlayer for analysis spot 1 would explain the lower Bi signal, despite it having a bigger Bi impurity based on the spot size analysis.

These findings obtained from varying the spot size at analysis spots 1 and 2 are summarized and represented schematically in Fig. 6. We show that the size of the X-ray beam spot can help estimate the bismuth cluster size. A drop in the Bi:S intensity ratio indicates that the X-ray spot size used must be greater than the Bi cluster size, and at X-ray spot sizes where the Bi:S ratio is plateaued, the beam spot must be fully within the Bi cluster. We also show that the thickness of the $MoS_2$ above a detected Bi impurity can be estimated with XPS. Furthermore, by considering this thickness, we can explain why the Bi:S ratio is higher for spot 2 at a spot size of 200 µm despite being the impurity being smaller than that in spot 1.

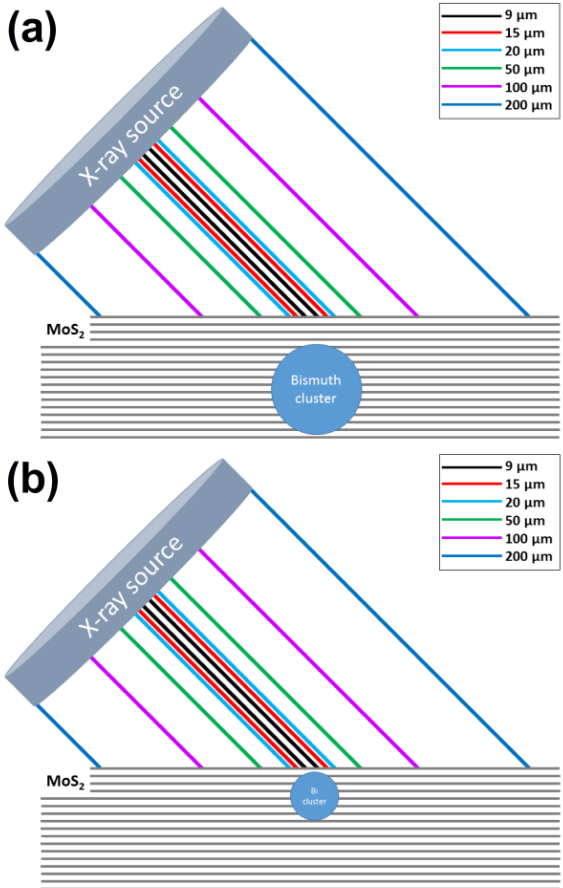

**Figure 6.** Diagrams representing the cluster size and MoS$_2$ overlayer thickness determined at analysis spot 1 (a) and analysis spot 2 (b).

### 3.4. Impurity Concentrations

The concentration of bismuth in an entire geological MoS$_2$ sample can be estimated based on our findings from spatially resolved XPS. Taking the area scanned for XPS mapping, 12 out of 81 laterally arranged spots were found to have Bi presence; that is, about 15% of that area contained some form of Bi impurities. Taking that area to be representative of a chunk of geological MoS$_2$, we can assume that an entire natural MoS$_2$ crystal is composed of approximately 15% Bi. Given that the Bi impurities are not homogenously distributed throughout the material, estimates from XPS are anticipated to be unreliable. Therefore we performed ICP-OES to determine the actual concentration of Bi.

ICP-OES composition measurements revealed Bi impurity levels to be ~100 ppmw. The entire specimen was dissolved in a 15 mL solution, indicating that ~20 wt% of the geological MoS$_2$ crystal was Bi. It can be seen that there is only a small difference between the ICP-OES measurement of ~20% Bi and the XPS measurement (from the XPS map) of ~15% Bi, indicating that the Bi content in the flakes analyzed were likely within this range.

### 3.5. Implications

These bismuth impurity compounds have consequences on the properties of the MoS$_2$. It would change the chemical bonding environment and electronic properties in its vicinity. However, since devices made using exfoliated flakes of MoS$_2$ are typically carried out after visual inspection focused on finding 'clean' flakes, it is likely that these micron-sized clusters of Bi are avoided. This is especially important when micrometer-sized contacts are used with MoS$_2$, which is typical in recent studies[2-4, 6, 31-40]. Because we know, through our spot size analysis, that the size of these bismuth impurities in MoS$_2$ is in the micrometer range, they could potentially dominate the properties of the MoS$_2$ if proper visual inspection and selection of good flakes is not performed. The presence of these micron-sized Bi clusters also highlights the difficulty of achieving large-area pristine MoS$_2$ without variability when using geological material.

We also know that these defects are present in clusters rather than uniform distributions, suggesting that the substitutional impurity concentration may be significantly lower than worst case estimates based on ICP-OES and ICP-MS[18] concentrations. The concentrations of substitutional Bi and smaller Bi clusters however, remain an unknown quantity that may be partly responsible for the atomic scale defects observed by Addou et al. and may therefore have a significant impact on device properties.

### 4. Conclusions

In this work Bi impurities in geological MoS$_2$ have been spatially resolved with XPS mapping. Bi impurities have previously been found in natural MoS$_2$ materials using mass spectroscopy techniques[18]. In this work, we show through XPS mapping that these impurities are localized in certain regions of the bulk MoS$_2$ crystal. Majority of the surface is chemically pristine MoS$_2$ as verified by XPS, but as shown in the concentration map, there are bismuth-rich areas. It was observed that the size of these impurities is in the order of tens of microns. Specific selected points indicate impurity sizes of somewhere between 15 to 50 μm. XPS also suggests that any Bi present in MoS$_2$ is not elemental Bi, but rather bismuth oxide or sulfate compounds (Bi$_2$O$_3$/Bi$_2$O$_5$/Bi$_2$(SO$_4$)$_3$). The abundance of these impurities in geological MoS$_2$ brings to light the importance of developing controlled synthesis and growth methods for MoS$_2$ as well as other TMDCs before they can be widely integrated into various device applications.

**Acknowledgements**

The XPS used for this work was provided through the NSF-MRI Award #1626201.

**Supplementary Material**

**MoS₂ Impurities: Chemical Identification and Spatial Resolution of Bismuth Impurities in Geological Material**

Maria Gabriela Sales[a], Lucas Herweyer[a], Elizabeth Opila[a], Stephen McDonnell[a,*]

[a]Department of Materials Science and Engineering, University of Virginia, Charlottesville, VA 22904
*Corresponding author: mcdonnell@virginia.edu

I.     **XPS of Geological MoS₂ from SPI Supplies**

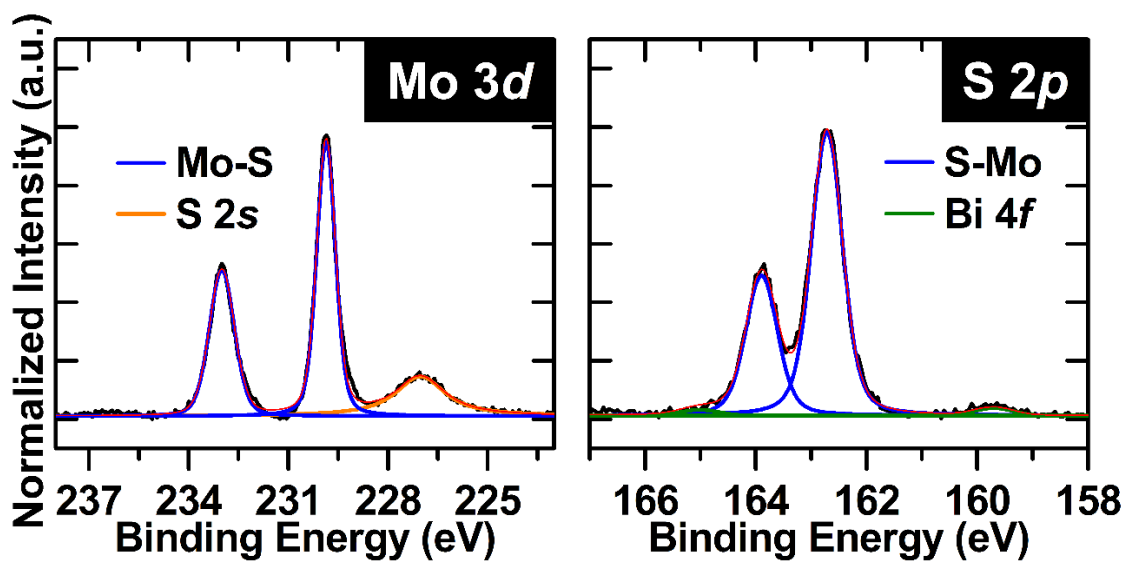

**Figure S1.** Representative XPS spectra of geological MoS₂ from SPI Supplies showing the presence of Bi impurities.

II.     **Details of ICP-OES Results**

**Table S1.** Results from ICP-OES measurements.

| Element | Bi | Mo | Re | S | Sb | W |
|---|---|---|---|---|---|---|
| ppmw | 102.04 | 173.32 | 1.51 | 19.18 | 1.38 | 0.23 |
| Weight (mg) | 1.5305 | 2.5999 | 0.0227 | 0.2876 | 0.0208 | 0.0034 |
| Wt% | 21.87% | 37.14% | 0.32% | 4.11% | 0.30% | 0.05% |

    We note that the sum of the weights obtained in the second row of Table S1 does not amount to 7 mg, which was the initial weight of the geological MoS₂ flake digested for ICP-OES (see Material and Methods). We speculate three possible reasons for this. First, the aqueous solution used for digestion could have reacted with the sample, possibly resulting in volatile species that escaped as gaseous products or the formation of precipitates that settled at the bottom of the solution. Second, spectral overlap in the

peaks that were analyzed could have potentially affected the calculations. Lastly, the ICP-MS study conducted by Addou et al. showed other elemental impurities in geological $MoS_2$, thus it is possible that other impurities present in the $MoS_2$ were not measured in our ICP-OES analysis. Aside from Bi, Mo, Re, S, Sb, and W, as shown in Table S1, we also checked for the presence of Ag, Ca, Cd, and Fe. We did not detect any Ag, Ca, Cd, and Fe through ICP-OES.

### III. Calculation of Bi Concentration Using Relative Sensitivity Factors (RSFs)

The Bi concentration in each of the 12 spots that showed Bi presence in the XPS map (Fig. 2) can be calculated based on the following equation, assuming only Mo, S, and Bi are detected in each spot:

$$Bi\ concentration\ (\%) = \frac{\frac{I_{Bi\ 4f}}{Bi\ 4f\ RSF}}{\frac{I_{Mo\ 3d}}{Mo\ 3d\ RSF} + \frac{I_{S\ 2p}}{S\ 2p\ RSF} + \frac{I_{Bi\ 4f}}{Bi\ 4f\ RSF}} \times 100$$

where $I_X$ is the intensity of the X core level peak and the PHI-recommended RSFs, corrected for the geometry of the PHI VersaProbe III system used, are as follows:
Mo $3d$ RSF = 87.535;  S $2p$ RSF = 17.497;  Bi $4f$ RSF = 244.962.

**Table S2.** Calculated Bi concentrations in the 12 spots with Bi presence (shown in Fig. 2) using Mo $3d$, S $2p$, and Bi $4f$ peak intensities and RSFs.

| Spot Number (based on Fig. 2a) | $I_{Mo\ 3d}$ | $I_{S\ 2p}$ | $I_{Bi\ 4f}$ | $\frac{I_{Mo\ 3d}}{Mo\ 3d\ RSF}$ | $\frac{I_{S\ 2p}}{S\ 2p\ RSF}$ | $\frac{I_{Bi\ 4f}}{Bi\ 4f\ RSF}$ | Bi concentration (%) |
|---|---|---|---|---|---|---|---|
| 2-14 | 2666.35 | 55.028 | 1095.119 | 30.46039 | 0.224639 | 62.58896 | **0.24** |
| 2-46 | 9250.7 | 236.952 | 3703.15 | 105.68 | 0.967301 | 211.6449 | **0.30** |
| 2-58 | 5358.58 | 81.0968 | 2109.257 | 61.21643 | 0.331059 | 120.5496 | **0.18** |
| 2-63 | 3520.93 | 1061.318 | 1238.07 | 40.22311 | 4.332582 | 70.75899 | **3.76** |
| 2-64 | 2889.64 | 1095.869 | 1246.344 | 33.01125 | 4.473629 | 71.23187 | **4.11** |
| 2-65 | 4794.892 | 41.8684 | 1952.562 | 54.77685 | 0.170918 | 111.5941 | **0.10** |
| 2-68 | 3396.29 | 27.0195 | 1393.24 | 38.79922 | 0.110301 | 79.62736 | **0.09** |
| 2-69 | 2982.23 | 286.911 | 1239.929 | 34.069 | 1.171247 | 70.86523 | **1.10** |
| 2-78 | 2710.75 | 44.0587 | 1024.592 | 30.96761 | 0.179859 | 58.55815 | **0.20** |
| 2-79 | 2371.819 | 70.7078 | 935.199 | 27.09566 | 0.288648 | 53.44911 | **0.36** |
| 2-80 | 1727.932 | 219.877 | 664.641 | 19.7399 | 0.897596 | 37.986 | **1.53** |
| 2-81 | 2109.75 | 268.854 | 767.898 | 24.10179 | 1.097533 | 43.88741 | **1.59** |

Averaging the last column over 81, which is the total number of spots analyzed in the XPS map, the estimated Bi concentration in the 800 µm x 800 µm area analyzed is approximately 0.17%. Note that the Bi concentration in all other spots not included in Table S2 were assumed to be 0%.